     \documentclass{article}       
     \usepackage{e}                
\usepackage[dvips]{graphicx}
\begin{document}
\newcommand{\ox}{ $\rightarrow$ }
\newcommand{\qqb}{$q \overline{q}$ }
\newcommand{\ttb}{$t \overline{t}$ }
\newcommand{\emiss}{$\not\hskip-5truedd E_{T}$ }
\newcommand{\ptmiss}{$\not\hskip-7truedd P_{T}$ }
\newcommand{\pT}{$p_{T}$ }
\newcommand{\X}{\times}
\newcommand{\lov}{\overline{l}}
\branch{C}   
%

\title{Production of Excited Neutrino at LHC}
\author{A. Belyaev$^{1}$\and C. Leroy\inst{2}\and R. Mehdiyev\inst{2,3}}
\institute{ 
 Department of Physics, Florida State University, Tallahassee, FL, USA.\and
Universit\'e de Montr\'eal, D\'epartement de Physique, Montr\'eal, H3C 3J7, Canada.\and
on leave of absence from Institute of Physics, Azerbaijan National Academy of Sciences, 
370143, Baku, Azerbaijan.}
\maketitle
\begin{abstract}
We study the potential of the CERN LHC in the search for the single
production of excited neutrino through gauge interactions.  
Subsequent decays of excited neutrino via gauge interactions are examined. 
The mass range accessible with the ATLAS detector is obtained. 
\end{abstract}
%
\vspace*{-8.5cm}
\begin{flushright}
{FSU-HEP-040110}
\end{flushright}
\vspace*{8.5cm}
\section{Introduction}
The proliferation of quarks and leptons can be naturally explained
by the assumption that they are composite  objects. According  to
models of compositeness\cite{1}, known fermions are bound states
of more fundamental constituents -- preons \cite{2}  or a fermion
and a boson \cite{3}. In the framework of these models, constituents of
known fermions interact by means of new strong gauge interactions.

One of the main  consequences of the non-trivial substructure of the 
standard model (SM) fermions would be a rich spectrum of excited  
states\cite{1,baur}. 
Observation of such fermionic excitations would be clear evidence
of the  underlying subtructure of known fermions. 
Therefore,  one of the  tasks
of great importance for TeV energy scale colliders is to probe
possible substructures of leptons and quarks and test the variety 
of compositeness models.

The SM can be considered as the low energy limit of a more
fundamental theory which is characterized by a large mass
scale $ \Lambda $.  The existence of four-fermion contact
interactions would be a signal of  new physics beyond the
SM. The nature of this new  physics can be probed if the experimental 
energy scale is high enough. It is expected
that the next generation of hadron  colliders like the LHC,
which will achieve very high centre of mass energies,  will
extend the search for composite states.  In particular,
contact interactions may be an important source for excited 
lepton production at the CERN LHC.

The excited states of the SM fermions
can interact via SM gauge field interactions and
also via new gauge strong interactions between preons. 
The later leads to effective contact
interactions between quarks and leptons and/or their excited
states in the low energy limit.

Many recent  experimental  studies have been devoted 
to the search for quark and lepton compositeness and excited states
at LEP(\cite{lep-exlep}),
HERA(\cite{hera-exlep}), and TEVATRON(\cite{tevatron-exlep}).
No signals from contact interactions and excited fermions have 
been found so far. Studies mentioned above
put  limits: i) on the compositeness scale
in the range of 2-8~TeV, depending on the type of the contact 
interactions, and ii) on the excited fermion mass up to the collider 
center-of-mass energy.

Based on previous studies ~\cite{lhc-compos1,baur,lhc-compos2,oscar,exe}, we 
expect that the LHC collider will put the most stringent
constraints on the composite models and/or the masses of excited fermions. 

This paper aims at a study of the potential of LHC collider in the search for
the excited neutrino production which has not been studied in details,
previously.
This work is a continuation of previous works devoted to the study
of  the excited quark and excited electron
production~\cite{lhc-compos2}.

The paper is organized as follows.
In Section~2 we discuss effective Lagrangians for models used for our study.
Section~3 presents details of our study and results,
while Section~4 outlines the conclusions.

\section{Physical setup}

For the sake of simplicity we limit the number of parameters in our
study and assume the most simple realization of a model where the
spin of the excited fermions is  $\frac{1}{2}$ and that they are
isospin $\frac{1}{2}$ partners (higher spin representations are
considered in \cite{hispin}, for example).

 We assume also that an excited fermion has acquired mass before 
 SU(2)$\otimes$U(1) symmetry breaking. Therefore, we consider 
 their left- and their right- components in isodoublets.
 For example, we have the following assignments for the first generation of fermions:
 \[
 l_L=\left(\begin{array}{c} \nu_e \\ {\rm e} \end{array} \right)_L \ , \ e_R 
 \ \ \ \ \ \ \ \ \ \ \ \ \
\ ;
 \ \
   l^*_L=\left(\begin{array}{c} \nu_{e}^* \\ {\rm e}^* \end{array} \right)_L \ , \
   l^*_R=\left(\begin{array}{c} \nu_{e}^* \\ {\rm e}^* \end{array} \right)_R \
 \]

 \[
  q_L=\left(\begin{array}{c} {\rm u} \\ {\rm d} \end{array} \right)_L \ ,
   q_R=\left(\begin{array}{c} {\rm u} \\ {\rm d} \end{array} \right)_R \ ; \ \
   q^*_L=\left(\begin{array}{c} {\rm u}^* \\ {\rm d}^* \end{array} \right)_L \ ,\
   q^*_R=\left(\begin{array}{c} {\rm u}^* \\ {\rm d}^* \end{array} \right)_R .
 \]

Let us note that in order to avoid conflict with precision measurements of 
anomalous magnetic moment of muon (g-2) and protect light fermions from large radiative 
corrections one should require a chiral form of interactions
of excited fermions with SM ones~\cite{renard}.

The couplings of excited fermions~($f^*=l^*,q^*$) to gauge bosons are vector like:
\begin{eqnarray}
\label{laggauge1}
&&L^{1 gauge}_{\rm eff}=
\bar{f^*}\gamma^\mu
(f_{s}g_{s}\frac{\mbox{\boldmath$l^{a}$}}{2}\mbox{\boldmath$G^{a}$}_{\mu}+
g\frac{\mbox{\boldmath$\tau$}}{2}\mbox{\boldmath$W$}_\mu +
g^{'}\frac{Y}{2}B_\mu)f^*,
\\
&&
\mbox{\hspace*{-1cm}while transitions between ordinary and excited fermions are
uniquely fixed by magnetic-moment type}
\nonumber
\\
&&\mbox{\hspace*{-1cm}gauge-invariant interactions~\cite{lagr}:}
\nonumber
\\
&&L^{2 gauge}_{\rm eff}=
\frac{1}{2{\Lambda}}\bar{f}^*_{\rm R}\sigma^{\mu\nu}
(f_{s}g_{s}\frac{\mbox{\boldmath$\lambda^{a}$}}{2}\mbox{\boldmath$G^{a}$}_{\mu\nu}+
fg\frac{\mbox{\boldmath$\tau$}}{2}\mbox{\boldmath$W$}_{\mu\nu} +
f^{'}g^{'}\frac{Y}{2}B_{\mu\nu})f_{\rm L} + {\rm h.c.},
\label{laggauge2}
\end{eqnarray}
where  
$\Lambda$ is the compositeness scale.
$\mbox{\boldmath$G^{a}$}_{\mu\nu}$,
$\mbox{\boldmath $W$}_{\mu\nu}$ and
$B_{\mu\nu}$ are SU(3), SU(2) and  U(1) tensors with the  coupling constants 
$g_{s}$, \ $g$ \ and \  $g^{'}$, respectively;
$Y$ is the weak hypercharge with $Y$ = (-1) and  (1/3) 
for leptons and quarks, respectively;
$f_s$, $f$ and $f'$ are parameters depending on the underlying dynamics.

\begin{figure}
{\par\centering 
{\includegraphics[height=4cm]{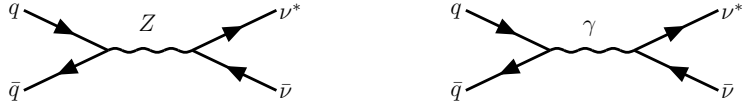}} \par}
\caption{Diagrams for single excited neutrino ($\nu^{\star}$) production via
photon and Z-boson exchange.} 
\label{diag_neut}
\end{figure}

\begin{figure}
{\par\centering 
{\includegraphics[height=4cm]{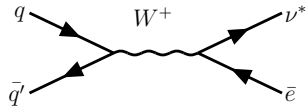}} \par}
\caption{Diagram for single excited neutrino ($\nu^{\star}$) production via
W-boson exchange.}
\label{diag_char}
\end{figure}

Lagrangian~\ref{laggauge2}
gives rise to the following $fermion-fermion^*-gauge\ boson$ vertices:
\begin{eqnarray}
\Gamma^{g\bar{f^*}f}_\mu&	=&\frac{g_s f_g}{2\Lambda}
q^\nu\sigma_{mu\nu}(1-\gamma^5)\\
\Gamma^{\gamma\bar{f^*}f}_\mu&	=&\frac{e}{2\Lambda} [e_f f'+T_3(f-f')]
q^\nu\sigma_{mu\nu}(1-\gamma^5)\\
\Gamma^{Z\bar{f^*}f}_\mu&	=&\frac{e}{2\Lambda}
\frac{I_3(c_w^2f+s_w^2f')-4e_f s_w^2f'}{s_w c_w}
q^\nu\sigma_{mu\nu}(1-\gamma^5)\\
\Gamma^{W\bar{f^*}f}_\mu&	=&\frac{e}{2\Lambda}
\frac{f}{\sqrt{2}s_w}
q^\nu\sigma_{mu\nu}(1-\gamma^5)
\end{eqnarray}
Excited fermions can be produced in pairs 
via interactions given by Eq.~(\ref{laggauge1})  as well as produced singly 
via interactions given by Eq.~(\ref{laggauge2}).
In this paper, we study single excited neutrino production
as the most promising reaction,  since it is less kinematically
suppressed compared to the case of production of the pair of excited neutrinos.
Excited neutrino can be singly produced at the LHC  according the process:
\begin{equation}
\label{eq_n}
q\bar{q}\to \nu\nu^{\star}
\end{equation}
i.e. via neutral current (see Fig.~\ref{diag_neut}) 
and  
via the charged current, in association with an electron (see Fig.~\ref{diag_char})
\begin{equation}
\label{eq_c}
q\bar{q'}\to e\nu^{\star}.
\end{equation}


\begin{figure}
{\par\centering 
{\includegraphics[width=9cm]{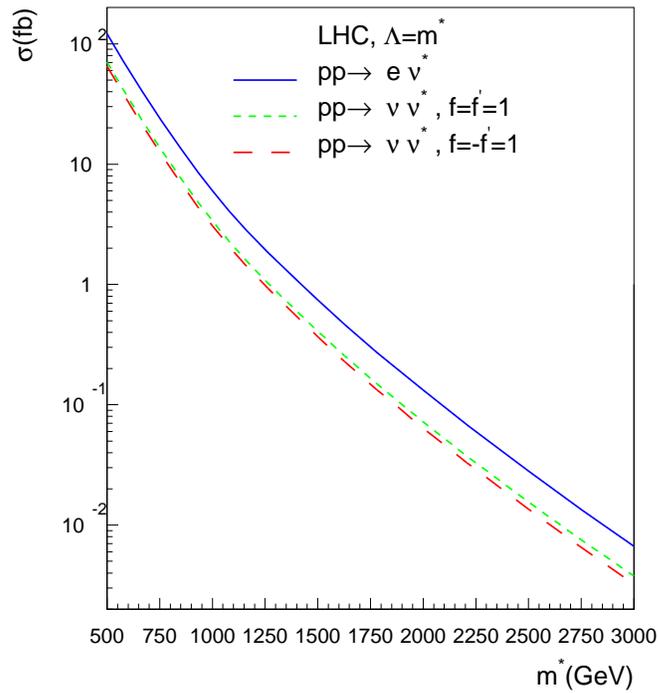}} \par}
\caption{
\label{cs-exnu}
Cross section of the single excited neutrino production versus the
excited neutrino mass, $m^{\star}$ at LHC for $\Lambda=m^{\star}$.
 Dashed and dotted line denote $f=f'=1$ and $f=-f'=1$
 choices, respectively, for the case of excited neutrino production via neutral currents.
 The cross sections shown account for the production of b
 oth excited neutrino and excited anti-neutrino at LHC.}
\label{cs_exnu}
\end{figure}

The couplings $f$ and $f'$ involved in the single excited neutrino production 
are not equal to each other, in general.
Therefore, the $\gamma\nu\nu^*$ coupling which is proportional to ($f-f'$)
can be non-vanishing. 
In Fig.~\ref{cs-exnu}, we present cross sections
for the processes~(\ref{eq_n}) and (\ref{eq_c}) as a function of the excited 
neutrino mass, 
$m^{\star}$ ($\Lambda=m^{\star}$) for two cases:  $f=f'=1$  and $f=-f'=1$.
The respective values for the cross sections are presented in 
Table~\ref{tab-cs}.
The cross section values were calculated using the CTEQ5L parton distribution 
function (PDF)\cite{12}. 
The QCD scale has been chosen equal to the excited neutrino mass.
We have checked that the systematical uncertainty due to the choice of others PDF
sets is about 20\%.

Excited neutrinos will decay to $\nu\gamma$, $\nu Z$ and $e W$
products, therefore, giving  rise to 
$\nu\nu\gamma$, $\nu\nu\nu$, $\nu l l$, $\nu q q$, $e\nu l$ and $e q q$
particles in the final state.
Branching ratios for excited neutrino decay which are defined
by gauge interactions and $f$ and $f'$ couplings
are presented in Table~\ref{tab-decay}.
One can see that for non-vanishing $\gamma\nu\nu^*$ couplings 
 $f=-f'=1$, the branching ratio, $Br(\nu^*\to \gamma \nu)$, is
of the order of 30\%, therefore the role of the 
$\nu\gamma$ channel would be significant in this case.
For the excited neutrino masses $m^{\star}>500$~GeV$>>M_Z,M_W$, 
the branching ratio of the excited neutrino decay 
does not depend on their masses (see Table~\ref{tab-decay}).

\begin{table}[h]
\caption{Cross sections (CompHEP) (in fb) for \qqb \ox $\nu^{\star}l$ and scale $\Lambda=m^{\star}$}
\centering{
\begin{tabular}{cc| ccccc}
\hline
$m^{\star}$(GeV) &
     &$  500  $&$ 1000 $&$     1500       $&$      2000      $&$     2500      $\\ \hline
pp \ox $e\nu^{\star}$    &   & $121.      $&$ 5.99 $&$ 7.43\X 10^{-1} $&$ 1.32\X 10^{-1} $&$ 2.82\X 10^{-2}$\\ 
pp \ox $\nu \nu^{\star}$ &
      $f=f^{\prime}=1$   &     $65.3      $&$ 3.07 $&$ 3.67\X 10^{-1} $&$ 6.40\X 10^{-2} $&$ 1.36\X 10^{-2}$\\
     &$f=-f^{\prime}=1$  &     $70.5      $&$ 3.37 $&$ 4.09\X 10^{-1} $&$ 7.21\X 10^{-2} $&$ 1.55\X 10^{-2}$\\
\hline
\end{tabular}}
\label{tab-cs}
\end{table}

\begin{table}[h]
\caption{Branching ratios(in \%) of excited neutrino decay via
         gauge interactions for  $\Lambda=m^{\star}$}
\centering{
\begin{tabular}{cc| cc}
\hline
\multicolumn{2}{c}{process} &\multicolumn{2}{|c}{for $m^{\star}$(GeV)} \\
          & & 500 & $>1000$        \\ \hline
$f=f^{\prime}=1$&
$\nu^*\to W  e     $& 61      & 61   \\
& $\nu^*\to Z \nu    $& 39      & 39   \\
&$\nu^*\to \gamma\nu$&  0      &  0   \\
\hline
$f=-f^{\prime}=1$&
$\nu^*\to W  e     $& 60      & 61   \\
&$\nu^*\to Z \nu    $& 12      & 12   \\
&$\nu^*\to \gamma\nu$& 28      & 27   \\
\hline
\end{tabular}}
\label{tab-decay}
\end{table}

\section{Simulations and results}

The simulations of excited lepton signal and relevant backgrounds
were  performed with COMPHEP\cite{16}, the COMPHEP-PYTHIA 
interface\cite{Belyaev:2000wn} and PYTHIA\cite{15} programs chain.
The ATLFAST\cite{17} code has been used to take into account the
experimental conditions prevailing at LHC for the  ATLAS detector.
The detector concept and its physics potential have been presented
in the ATLAS Technical Proposal\cite{18} and the ATLAS Technical
Design Report\cite{19}.  The ATLFAST program for fast detector
simulations accounts for most of the detector features: jet
reconstruction in the calorimeters, momentum/energy smearing for
leptons and photons, magnetic field effects and missing transverse
energy. It provides a list of reconstructed  jets, isolated
leptons and photons. In most cases, the detector dependent
parameters were tuned to values expected for the performance of
the ATLAS detector obtained from full simulation.

The electromagnetic calorimeters were used to reconstruct the energy of leptons
in cells of dimensions $ \Delta \eta \X \Delta \phi =0.025\X 0.025 $
within the pseudorapidity range $ -2.5<\eta <2.5 $; $\phi$ is the azimuthal angle. 
The electromagnetic energy resolution is given by $ 0.1/\sqrt{E}(GeV)\bigoplus 0.007 $ 
over this pseudorapidity ($ \eta $) region.
The electromagnetic showers are identified as leptons when they lie within a cone
of radius $ \Delta R=\sqrt{(\Delta \eta)^{2} \X (\Delta \phi)^{2}} = 0.2 $ 
and possess a transverse energy $ E_{T}>5 $ GeV. 
Lepton isolation criteria were applied, requiring a distance $\Delta R>0.4$
from other clusters and maximum transverse energy deposition, $ E_{T}<10 $ GeV, 
in cells in a cone of radius $ \Delta R=0.2 $ around the direction of lepton emission.

It must be mentioned that standard parametrization in the ATLFAST has been
used for the leptonic resolution but detailed studies are needed, using
test beam  data and GEANT full simulation to validate the extrapolation of
the resolution function to leptonic energies in the TeV range.

\subsection{\qqb $\rightarrow \nu^{\star} \nu$ subprocess}

\begin{figure}
\centerline{\includegraphics[height=3cm]{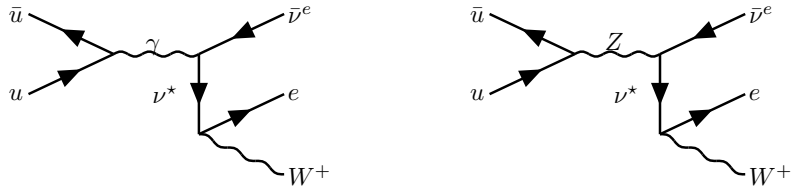}}
\caption{Production and decay of excited neutrino ($\nu^{\star}$) to $W$ and electron ($e$).} 
\label{fig1}
\end{figure}

For this type of subprocess, we consider only the decay of excited neutrino to a $W$ and
an electron mediated by gauge interactions. Typical Feynman diagrams relevant for this
process are shown in Fig.~\ref{fig1}.

For the $W$ decay, we limited ourselves to the case $W$ \ox jets. In the case of
semileptonic decays of $W$, the final state consists of two neutrinos, giving a large
uncertainty in the excited neutrino mass reconstruction.  

The signal signature for the selected reaction ($\nu^{\star} \rightarrow We$) consists
of an electron and two jets. 

We considered three SM backgrounds:

\begin{itemize}
\item  \ttb pair production, where one $W$ decays to jets and 
the second $W$ decays into an electron and a neutrino.
\item $WW$ pair production with the same decays as above.
\item  $W + jets$ production, where $W$ is allowed to decay to an electron and a 
neutrino.
\end{itemize}

The following cuts were used to separate the signal from background:

\begin{itemize}

\item The electron was required to be emitted in the pseudorapidity region $|\eta |<2.5$
and its transverse momentum was required to be at least 150 ( 250, 300 ) GeV for
$m^{\star}$ masses of 500, ( 750, 1000 ) GeV, respectively. 

\item The transverse momenta of two jets were required to be at least 50 GeV.
\item It was required to have $W$ mass reconstructed with two jets in the 
$60 - 100$ GeV mass window (mainly to suppress the dominant $W+jets$ background).  
\item The missing transverse momentum cut, \ptmiss, was required to be at least 300 GeV. 
 \end{itemize} 

\begin{figure}[ht]
\begin{center}
\begin{tabular}{c c}
\includegraphics[width=8cm,height=8cm]{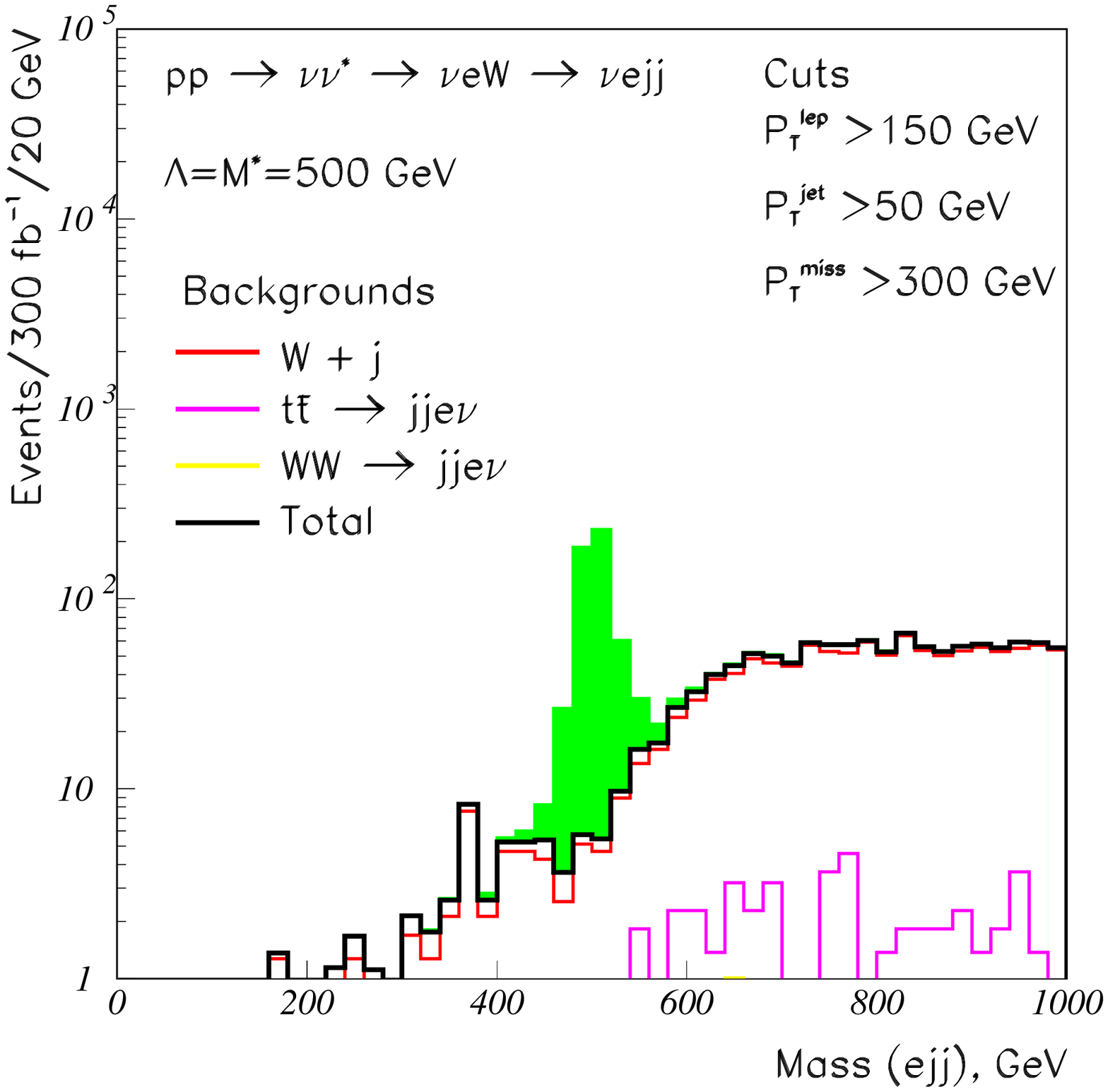}
\includegraphics[width=8cm,height=8cm]{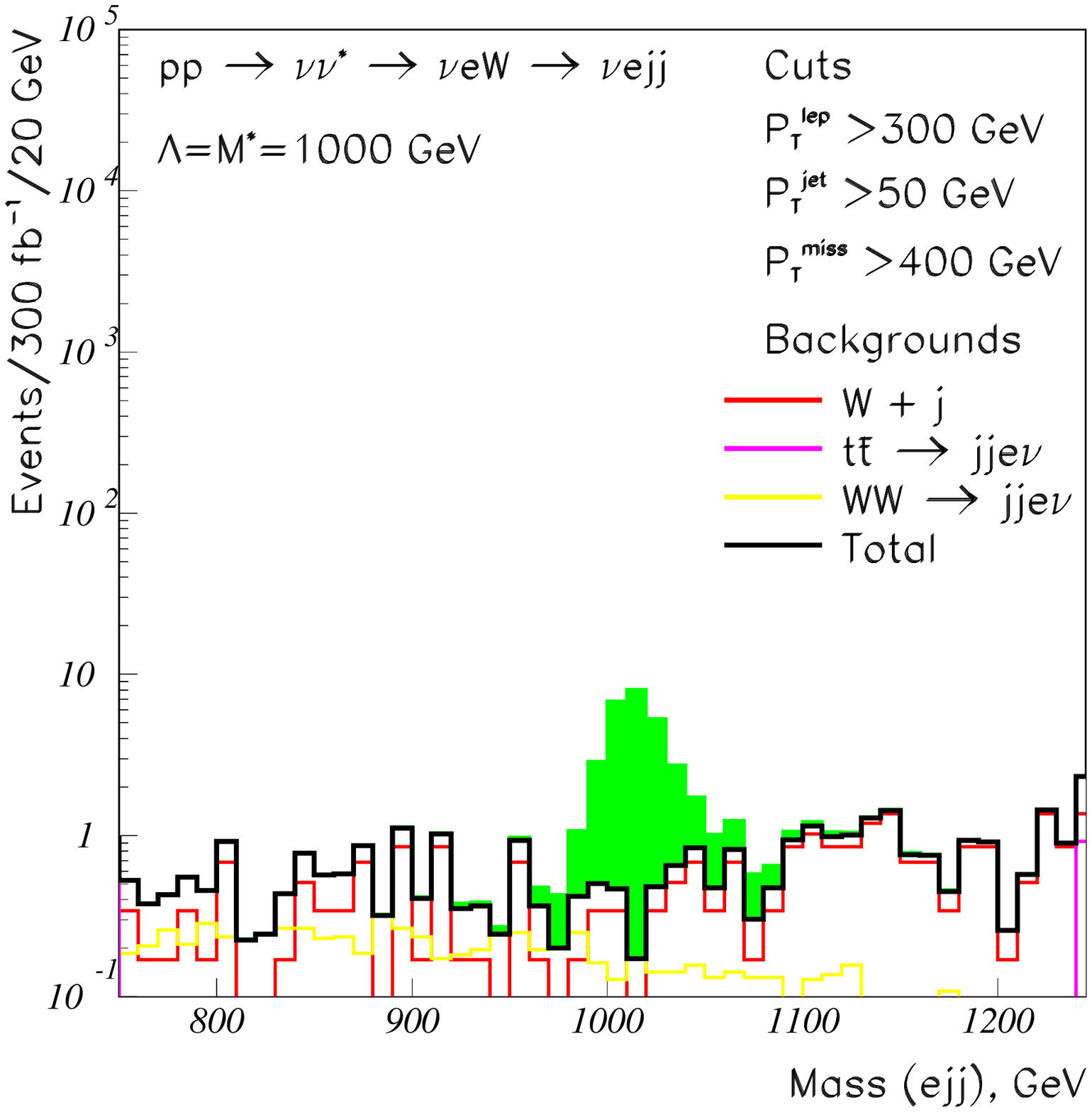}
\end{tabular}
\end{center}
\caption{Invariant mass distributions of $\nu^{\star}$ ($\rightarrow We$) and $W$'s  
decay to jets for $m^{\star}=500$ GeV (left) and $m^{\star}=1000$ GeV (right). 
The integrated luminosity is $300 fb^{-1}$.} 
\label{fig2}
\end{figure}

The resulting invariant mass distributions of the (electron-jet-jet) system are presented in
Fig.~\ref{fig2} for the mass of the excited neutrino $m^{\star}=500$ GeV (left) and
$m^{\star}=1000$ GeV (right).
 
The resonances are clearly seen above the total background.
The distributions were normalized to an integrated luminosity of $L=300$ fb$^{-1}$.

\subsection{\qqb $\rightarrow e \nu^{\star}$ subprocess}

\subsubsection{$\nu^{\star} \rightarrow We$ decay channel.}

\begin{figure}
\centerline{\includegraphics[height=3cm]{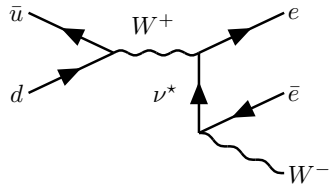}}
\caption{Production and decay of excited neutrino ($\nu^{\star}$) to $W$ and electron.} 
\label{fig3}
\end{figure}

Here again we considered the decay of excited neutrino to $We$ mediated by gauge interactions (Fig.~\ref{fig3}).

We have studied cases with $W$ decay to jets, as well as, with semileptonic decays of $W$ into an electron and a neutrino.

In the case of $W$ decay to $e \nu$, the final state consists of three electrons and a neutrino. The signal signature is sought in the system of  two electrons and a neutrino.

For this final state we studied two SM backgrounds:

\begin{itemize}
\item $W+Z$ production where $W$ decays to $e \nu$ and $Z$ decays to electron - positron pair.   
\item  $WWW$ production, where all three $W$ allowed to decay only into an electron and a neutrino.
\end{itemize}

The following cuts were used to separate the signal from background:

\begin{itemize}
\item The transverse momenta of three electron were required to be at least 50 GeV,
and to be within the pseudorapidity acceptance of $|\eta |<2.5$. 
\item It was required to have $W$ \ox $e \nu$ mass reconstructed in the 
($70 - 90$) GeV mass window. This requirement was used as a constraint for obtaining  
the energy of the neutrino.
\item The reconstruction of the two electron masses in the ($80 - 100$) GeV mass window 
has been used as a $Z$ production veto to suppress the $W+Z$ background. 
\end{itemize}

In the case of $W$ decays to jets, the final state consists of two electrons and two jets. For this reaction the signal signature consists of an electron and two accompanying jets. 

For the relevant SM backgrounds corresponding to this final state we used:

\begin{itemize}
\item $Z+jets$ production where $Z$ decays to an electron - positron pair.   
\item  $WWW$ production, where two $W$'s are allowed to decay into an electron and a 
neutrino and the third $W$ to jets. 
\end{itemize}

For this signal signature, the corresponding cuts used were:

\begin{itemize}
\item The transverse momenta of two electrons were required to be at least 150 GeV,
and to be within the pseudorapidity region of $|\eta |<2.5$. 
\item The transverse momenta of two jets were required to be at least 20 GeV.
\item It was required to have the mass of $W$ \ox jets reconstructed in the 
($70 - 90$) GeV mass window, for the suppression of the dominant $Z+jets$ background.
\item The reconstruction of the two electron masses in the ($80 - 100$) GeV mass window also has been used as a veto to suppress the $Z+jets$ background. 
\end{itemize}

\begin{figure}[ht]
\begin{center}
\begin{tabular}{c c}
\includegraphics[width=8cm,height=8cm]{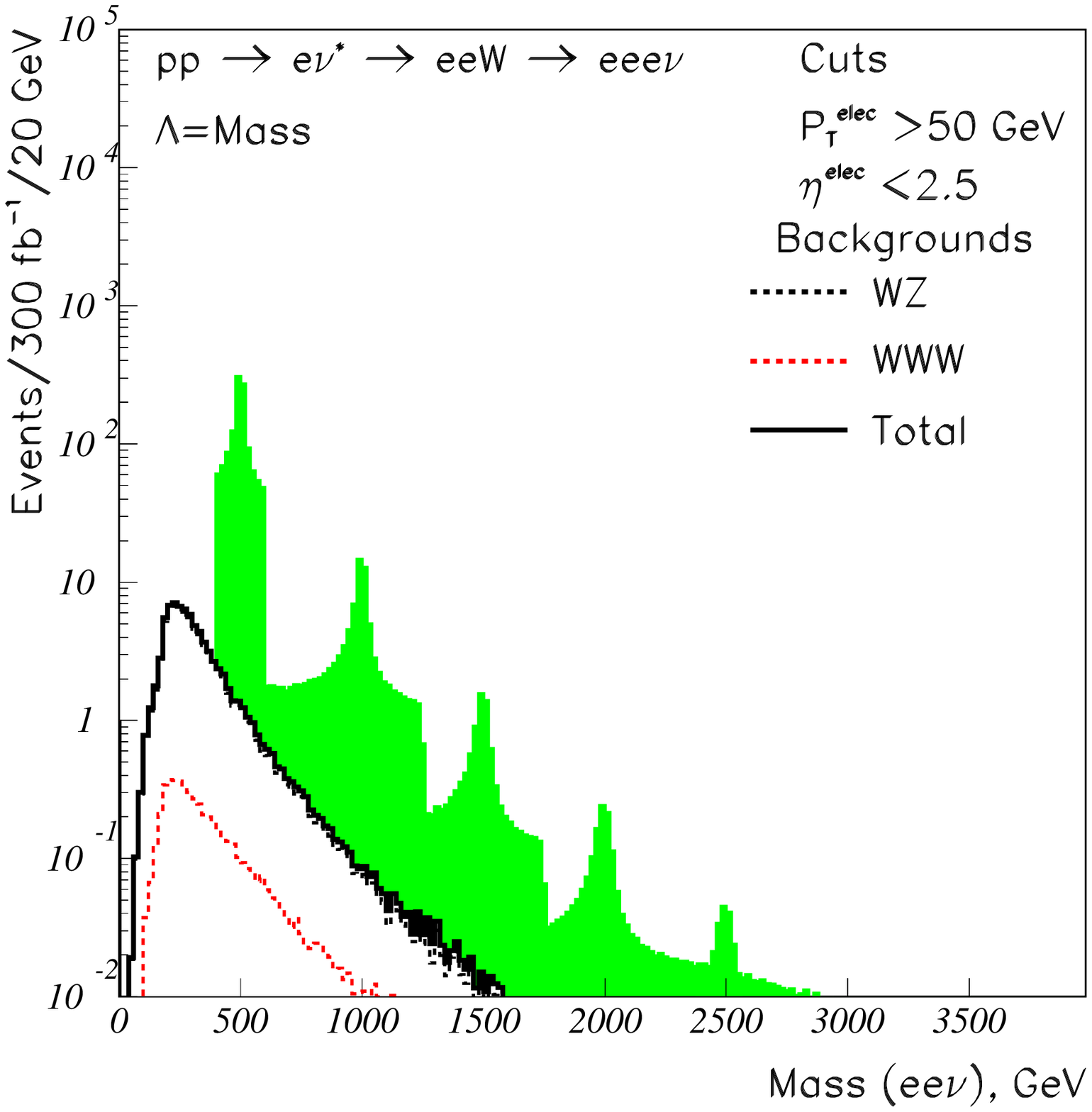}
\includegraphics[width=8cm,height=8cm]{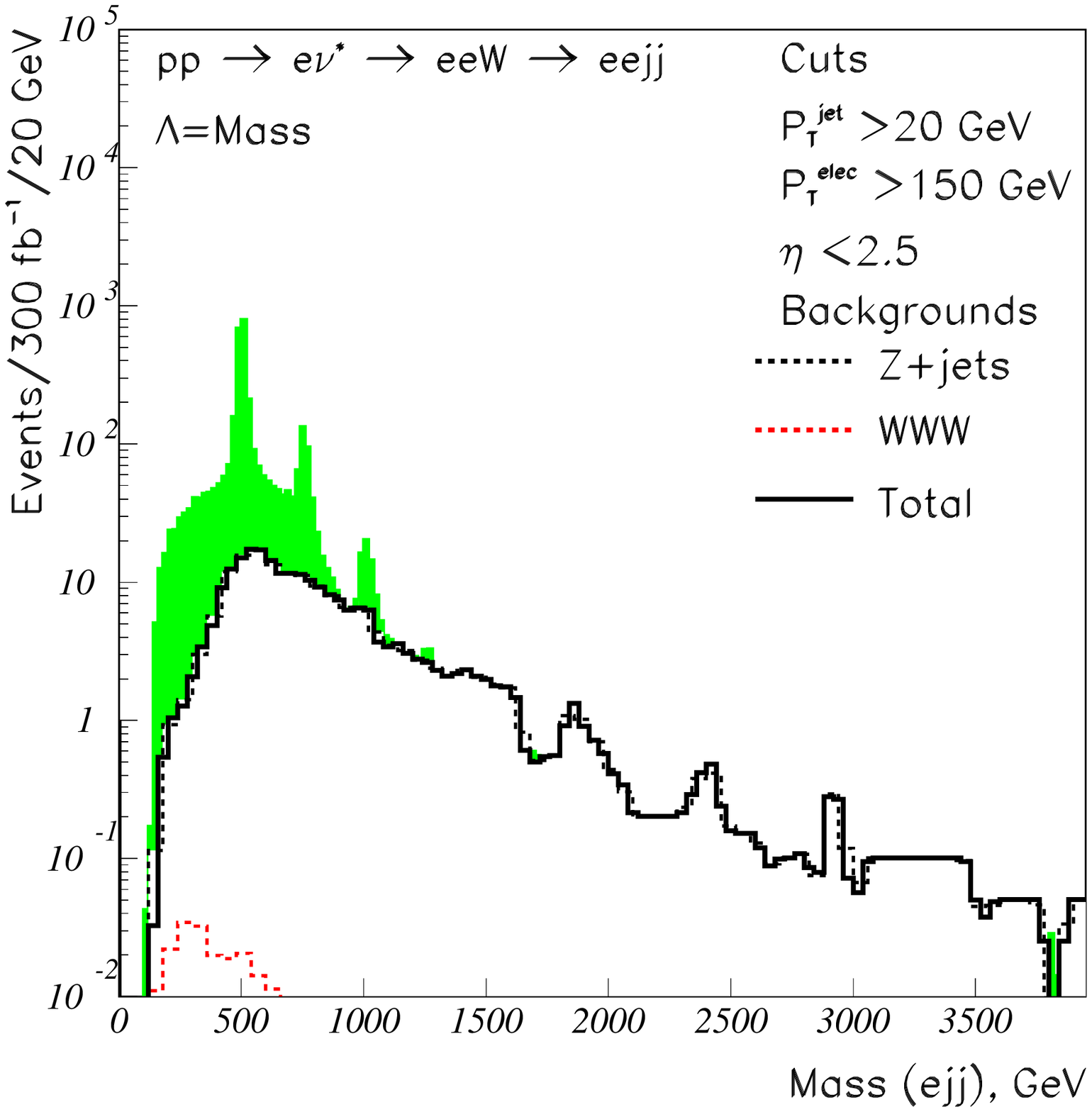}
\end{tabular}
\end{center}
\caption{Invariant mass distributions of $\nu^{\star}$ ($\rightarrow We$) for the $W$  
decay mode to $e\nu$ (left) and to jets (right). The integrated luminosity is $300 fb^{-1}$.} 
\label{fig4}
\end{figure}

The resulting invariant mass distribution is presented in Fig.~\ref{fig4} for the system of two electrons and neutrino (left side) and for the system of electron-jet-jet (right side) for different masses of the excited neutrino.
The distributions were normalized to an integrated luminosity of $L=300$ fb$^{-1}$.

\subsubsection{$\nu^{\star} \rightarrow Z\nu$ decay channel.}

The decay of the excited neutrino to $Z\nu$, mediated by gauge interactions, was considered in this section. 

\begin{figure}[ht]
\centerline{\includegraphics[height=3cm]{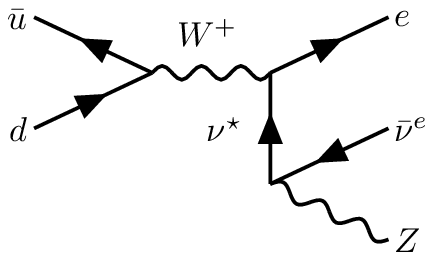}}
\caption{Production and decay of the excited neutrino ($\nu^{\star}$) to $Z$ and $\nu$.} 
\label{fig5}
\end{figure}

The corresponding diagram is presented in Fig.~\ref{fig5}.

For the $Z$ decay, we studied the cases where $Z$ decays to jets or $Z$ decays into $\mu^{+}\mu^{-}$ pairs.

In the case of $Z$ decays to jets, the final state consists of an electron, a neutrino and two jets. The signal sought for an energetic electron with two accompanying jets in the presence of a large missing transverse energy.

For this signal we considered three SM background:

\begin{itemize}
\item  $W + jets$ production, where $W$ are allowed to decay into an electron and a neutrino.
\item  \ttb pair production, where one $W$ decays to jets and the second $W$ into an electron and a neutrino.
\item $WW$ pair production with the same decays as above.
\end{itemize}

The following cuts were used to separate the signal from backgrounds:

\begin{itemize}
\item The transverse momentum of an electron was required to be at least 170 (200, 400) GeV for excited neutrino masses of 500 (750, 1000) GeV
and to be emitted within pseudorapidity acceptance of $|\eta |<2.5$. 
\item The transverse momenta of two jets were required to be at least 40 GeV.
\item It was required to have $Z$ \ox jets mass reconstructed in the 
($80 - 100$) GeV mass window to suppress the dominant $W+jets$ background.
\item The missing transverse momentum cut, \ptmiss, was required to be at least 400 GeV.
\end{itemize}

\begin{figure}
\begin{center}
\begin{tabular}{c c}
\includegraphics[width=8cm,height=8cm]{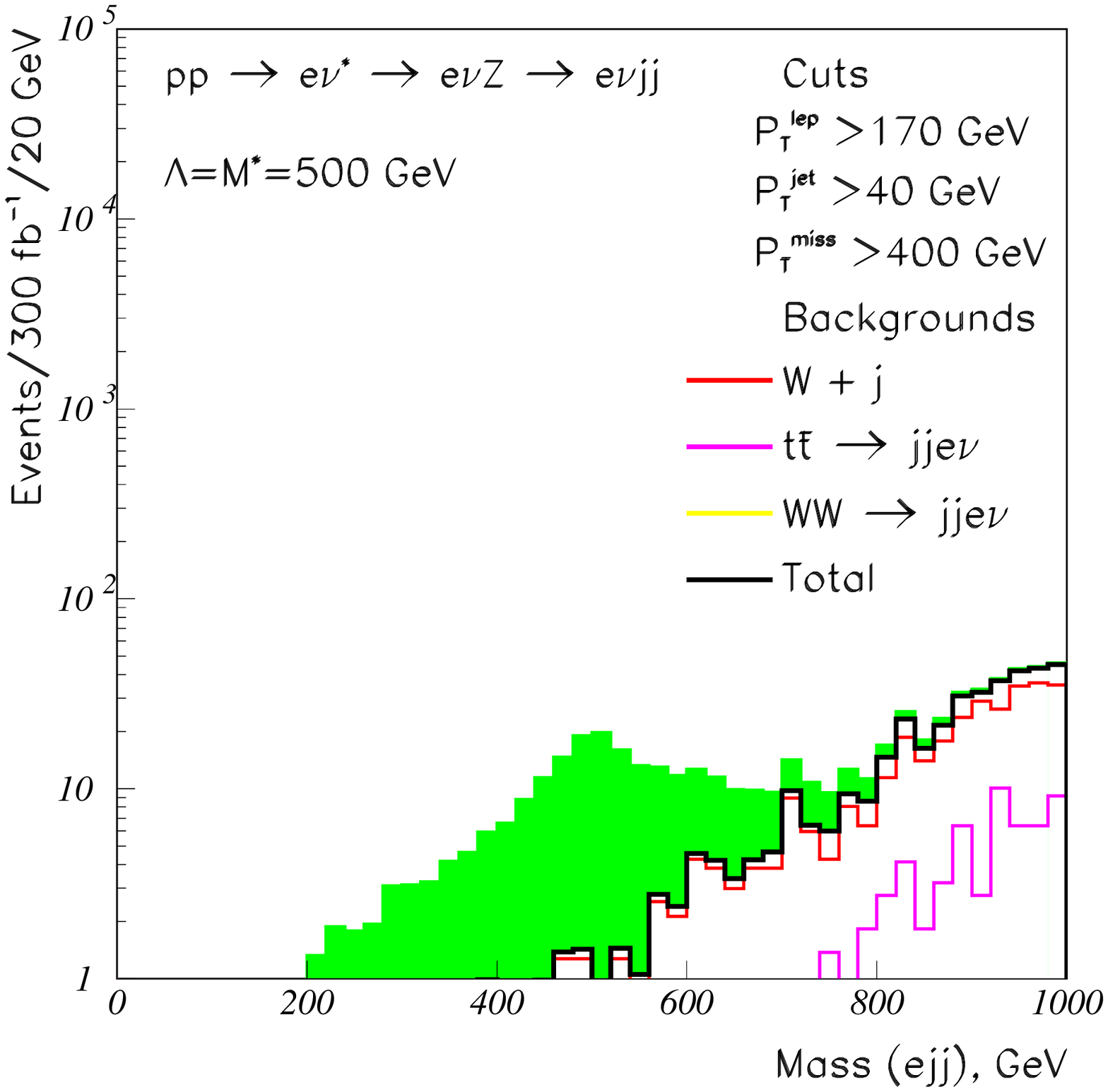}
\includegraphics[width=8cm,height=8cm]{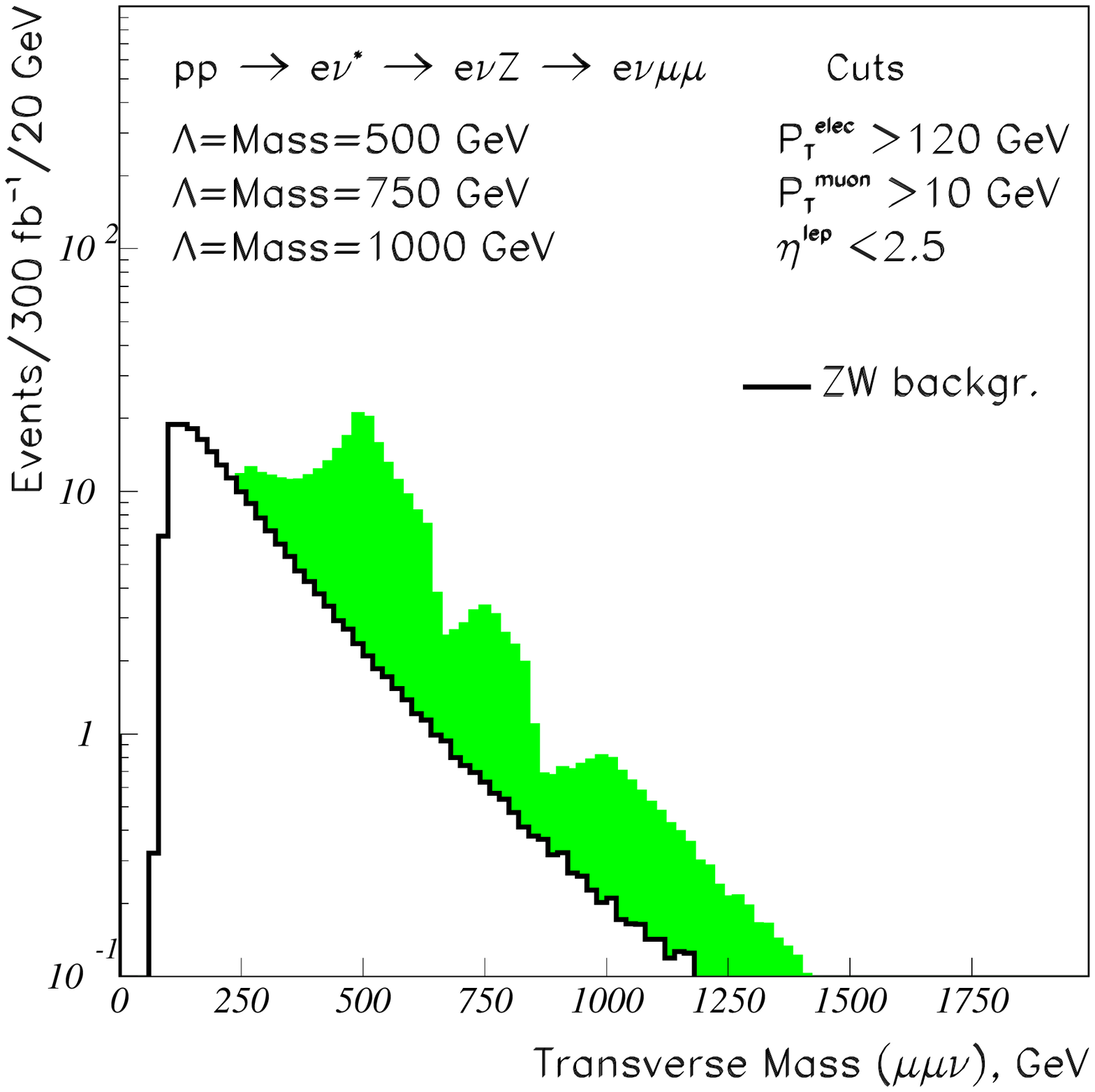}
\end{tabular}
\end{center}
\caption{Invariant mass distributions of $\nu^{\star}$ ($\rightarrow Ze$) for $Z$  
decay mode to jets for $m^{\star}=500$ GeV (left) and $\nu^{\star}$ ($\rightarrow Ze$) for $Z$  decay mode to $\mu^{+}\mu^{-}$ (right). 
The integrated luminosity is $300 fb^{-1}$.} 
\label{fig6}
\end{figure}

The resulting invariant mass distributions of the electron-jet-jet system are presented in
Fig.~\ref{fig6} for the mass of the excited neutrino $m^{\star}=500$ GeV (left).
The distribution was normalized to an integrated luminosity of $L=300$ fb$^{-1}$.

For the case where $Z$ decays into $\mu^{+}\mu^{-}$ pairs, the final state consists of
two muons, an electron and a neutrino. The signal signature for this subprocess is
two muons and the missing transverse energy, accompanied by an energetic electron.

The natural SM background for this subprocess is the $W+Z$ production where $W$ decays to $e \nu$ and $Z$ decays into muons.

The following cuts were used to separate the signal from the background:

\begin{itemize}
\item The transverse momentum of an electron was required to be at least 120 GeV
within the pseudorapidity of $|\eta |<2.5$. 
\item The transverse momenta of two muons were required to be at least 10 GeV.
\item The missing transverse momentum cut, \ptmiss, was required to be at least 100 GeV. 
\end{itemize}

The resulting invariant mass distribution of two muons combined with the  missing transverse energy  is presented in Fig.~\ref{fig6} (right) for an integrated luminosity of $L=300$ fb$^{-1}$.

\subsubsection{$\nu^{\star} \rightarrow  \nu \gamma$ decay channel.}

\begin{figure}
\centerline{\includegraphics[height=3cm]{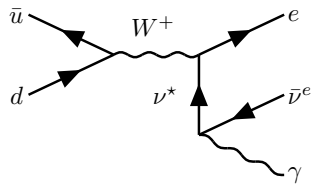}}
\caption{Production and decay of excited neutrino ($\nu^{\star}$) to $\gamma$ and $e$} 
\label{fig8}
\end{figure}

Another interesting subprocess is the decay of an excited neutrino to $\nu$ and a photon (Fig.~\ref{fig8}).

The signal consists of a photon and a neutrino in the presence of an energetic electron.

The natural SM background for this subprocess is the $W+\gamma$ production where $W$ decays to $e$ and $\nu$.

The cuts used to separate the signal from background are:

\begin{itemize}
\item The transverse momenta of an electron and a photon were required to be at least 50 GeV, and to be within the pseudorapidity acceptance of $|\eta |<2.5$. 
\end{itemize}

\begin{figure}
\begin{center}
\begin{tabular}{c}
\includegraphics[width=8cm,height=8cm]{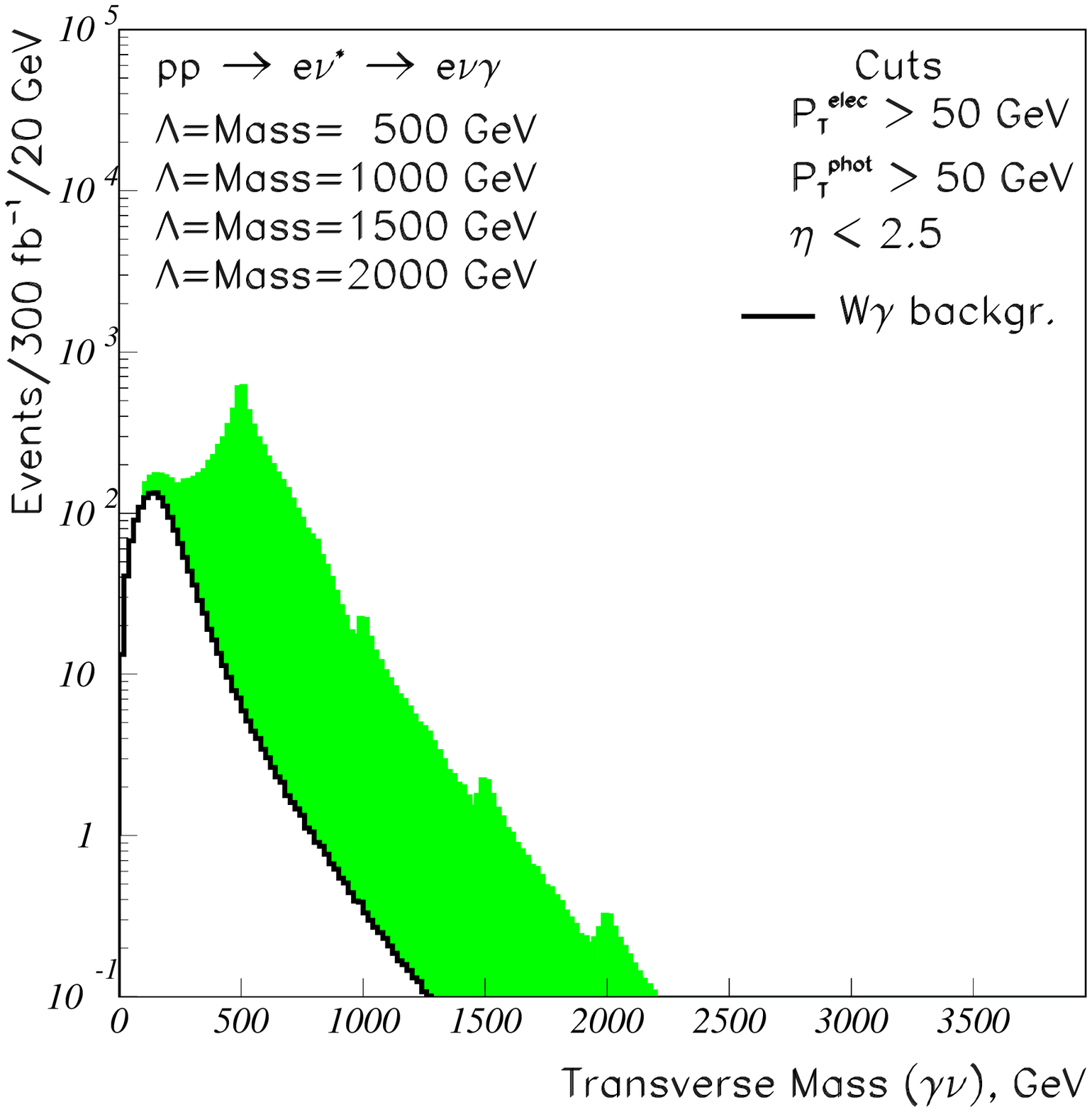}
\end{tabular}
\end{center}
\caption{Transverse mass distribution of $\nu^{\star} \rightarrow \nu \gamma$. The integrated luminosity is $300 fb^{-1}$.} 
\label{fig7}
\end{figure}

The resulting transverse mass distribution of the electron and missing transverse momentum is presented in Fig.~\ref{fig7} (right) for an integrated luminosity of $L=300$ fb$^{-1}$.

In Table~\ref{table3}, the corresponding signal significances, obtained for all studied subprocesses, are presented for an integrated luminosity of $L=300 fb^{-1}$.
The number of accepted signal and
background events were defined in the selected mass bin width ($\Delta M$).
For the mass bin width, the value was taken equal to a $\pm 2\sigma$ width of the 
invariant mass distribution around the excited neutrino peak's position.

\begin{center}
\begin{table}[ht]
\caption{The signal significances, $S/\sqrt{S+B}$,
$S$ for signal, $B$ for total background, and number of events are
calculated for an integrated luminosity of $L=300 fb^{-1}$,
$\Lambda = m^{\star}$ and various couplings within selected mass bin width ($\Delta M$).}
\centering{
\begin{tabular}{llllll} \hline
$m^{\star} (GeV)\rightarrow$   &   500  &  750  &  1000  &  1250 &  1500 \\ \hline
\hline \qqb \ox $\nu\nu^{\star}$ \ox $\nu eW$ \ox $\nu ejj$, $f=\pm f^{\prime}=1$ & & & &\\ 
\hline $\Delta M$, GeV         &    28  &   72   &   92   &  116 &   -   \\ 
\hline $S$                     &   472  &  103   &   24   &   5  &   -   \\
\hline $S/\sqrt{S+B}$          &    21  &   10   &    5   &   2  &   -   \\ \hline

\hline \qqb \ox $e\nu^{\star}$ \ox $eeW$ \ox $eee\nu$, $f=\pm f^{\prime}=1$ & & & & \\ 
\hline $\Delta M$, GeV         &    32  &   98   &  100  &  102  &  132 \\ 
\hline $S$                     &  1097  &  271   &   60  &   26  &    6 \\
\hline $S/\sqrt{S+B}$          &    33  &   16   &    7  &    5  &  2.2 \\ \hline

\hline \qqb \ox $e\nu^{\star}  $ \ox $eeW$ \ox $ee jj$, $f=\pm f^{\prime}=1$ & & & & \\ 
\hline $\Delta M$, GeV         &    36  &   92   &  120  &  160  &  180  \\ 
\hline $S$                     &  2015  &  488   &  108  &   23  &    3  \\
\hline $S/\sqrt{S+B}$          &    44  &   20   &    9  &    3  &   1.5 \\ \hline

\hline \qqb \ox $e\nu^{\star}  $ \ox $e\nu Z$ \ox $e\nu \mu\mu$, $f=f^{\prime}=1$&&&&\\ 
\hline $\Delta M$, GeV         &   100  &  260   &  400  &   -   &  -   \\
\hline $S$                     &   123  &   20   &   4   &   -   &  -   \\
\hline $S/\sqrt{S+B}$          &    10  &    4   &   2   &   -   &  -   \\  \hline

\hline \qqb \ox $e\nu^{\star}  $ \ox $e\nu Z$ \ox $e\nu jj$, $f=f^{\prime}=1$&&&&\\ 
\hline $\Delta M$, GeV         &   160  &  240   &  320  &  -  &  -   \\
\hline $S$                     &   141  &   36   &    6  &  -  &  -   \\
\hline $S/\sqrt{S+B}$          &    29  &    5   &    2  &  -  &  -   \\  \hline

\hline \qqb \ox $e\nu^{\star}  $ \ox $\gamma e \nu$, $f=-f^{\prime}=1$ &&&&\\ 
\hline $\Delta M$, GeV         &  120   &   160  &  200  & 204   &  240  \\
\hline $S$                     &  4218  &   546  &  196  &  63   &   27  \\
\hline $S/\sqrt{S+B}$          &    64  &    23  &   14  &   8   &    5  \\ \hline
\hline
\end{tabular}}
\label{table3}
\end{table}
\end{center}

As can be seen from Table~\ref{table3},  the highest reach for excited
neutrino production would be available in decays of the excited neutrino
to $\nu$ and a photon (due to a low background level ) for a non-zero 
$\nu\nu^*\gamma$ coupling at $f=-f^{\prime}=1$
but excited neutrino decay channel involving $W$ is also promising.
In case of  $f=f^{\prime}=1$, the $\nu\nu^*\gamma$ coupling vanishes, 
and excited neutrino decay channel involving $W$ becomes clearly dominant.  
The mass reach for the decay channel of neutrino with ($eee\nu$) final state 
is around 1500 GeV, practically independent from $f=f^{\prime}=1$ or $f=-f^{\prime}=1$
couplings choice.
At lower values of excited neutrino masses the signature with ($eejj$) in
the final state is more promising due to a better statistical
significance.  

Excited neutrino decay channels involving $Z$ bosons, due
to the smaller branching ratio, could be used only to confirm excited
neutrino observation, obtained from other channels. 
This channel could be observable
for the $f=f^{\prime}=1$ case and $m_{\nu^*}$ only below 1~TeV.
The case of  $f=-f^{\prime}=1$ is even less promising and it is not
presented in Table~\ref{table3} --- the number of signal events
goes down with about factor of 3.5 according to the $\nu^*\to Z\nu$ branching ratio
(see Table~\ref{tab-decay}).

\section{Conclusions}

Based on the prediction of a composite model of quarks and leptons, excited neutrino will be
possibly observed at CERN LHC. We have presented the results of excited single neutrinos
production and their subsequent decays through gauge interactions.   Rather clean signatures
are expected to be found for certain decays with neutrino in the final state. 
We have been studied two cases of $f,f'$ parameters:
the case of $f=-f^{\prime}=1$ which gives rise to a non-vanishing $\nu\nu^*\gamma$ 
coupling and the case of $f=f^{\prime}=1$.
For $f=-f^{\prime}=1$,
the highest reach is expected for $e \nu \gamma$ final state,
while in the case of $f=f^{\prime}=1$, $eee \nu$ and $eejj$ final states look most 
promising to reach large excited neutrino masses.   
We have found that singly produced excited neutrinos could be accessible up to a mass 
of 1.5 TeV at LHC,  assuming an integrated luminosity of $L=300fb^{-1}$.

\section{Acknowledgements}
This work has been performed within the ATLAS Collaboration with the help of the simulation framework and tools which are the result of the collaboration-wide efforts.
We would like also to thank G. Azuelos, H.Baer, O.\c{C}ak\i r, D. Froidevaux, F. Gianotti, I. Hinchliffe, L. Poggioli, L.Reina for their comments about the subject. C.L. and R.M. thank NSERC/Canada for their support.
This research was partially supported by the U.S. Department of Energy under contracts 
number DE-FG02-97ER41022 and DE-FG03-94ER40833.
\end{document}